\documentclass[authorversion, screen]{acmart}
\AtBeginDocument{%
  }

\setcopyright{acmlicensed}
\copyrightyear{2025}
\acmYear{2025}
\acmDOI{XXXXXXX.XXXXXXX}

\begin{document}

\title{Why Slop Matters}

\author{Cody Kommers}
\authornote{Corresponding Author}
\email{ckommers@turing.ac.uk}
\orcid{0009-0007-8985-0085}
\affiliation{%
  \institution{The Alan Turing Institute}
  \city{London}
  \country{UK}
}

\author{Eamon Duede}
\affiliation{%
  \institution{Purdue University}
  \country{USA}}

\author{Julia Gordon}
\affiliation{%
  \institution{Duke University}
  \country{USA}}

\author{Ari Holtzman}
\affiliation{%
  \institution{University of Chicago}
  \country{USA}}

\author{Tess McNulty}
\affiliation{%
  \institution{University of Illinois Urbana-Champaign}
  \country{USA}}

\author{Spencer Stewart}
\affiliation{%
  \institution{Purdue University}
  \country{USA}}

\author{Lindsay Thomas}
\affiliation{%
  \institution{Cornell University}
  \country{USA}}

\author{Richard Jean So}
\authornote{Equal senior author role.}
\email{richard.so@duke.edu}
\affiliation{%
  \institution{Duke University}
  \country{USA}}
\author{Hoyt Long}
\authornotemark[2]
\email{hoytlong@chicago.edu}
\affiliation{%
  \institution{University of Chicago}
  \country{USA}}


\renewcommand{\shortauthors}{Kommers et al.}

\begin{abstract}
  AI-generated "slop" is often seen as digital pollution. We argue that this dismissal of the topic risks missing important aspects of AI Slop that deserve rigorous study. AI Slop serves a social function: it offers a supply-side solution to a variety of problems in cultural and economic demand---that, collectively, people want more content than humans can supply. We also argue that AI Slop is not mere digital detritus but has its own aesthetic value. Like other "low" cultural forms initially dismissed by critics, it nonetheless offers a legitimate means of collective sense-making, with the potential to express meaning and identity. We identify three key features of family resemblance for prototypical AI Slop: superficial competence (its veneer of quality is belied by a deeper lack of substance), asymmetry effort (it takes vastly less effort to generate than would be the case without AI), and mass producibility (it is part of a digital ecosystem of widespread generation and consumption). While AI Slop is heterogeneous and depends crucially on its medium, it tends to vary across three dimensions: instrumental utility, personalization, and surrealism. AI Slop will be an increasingly prolific and impactful part of our creative, information, and cultural economies; we should take it seriously as an object of study in its own right.
\end{abstract}

\ccsdesc[500]{Applied computing~Media arts}
\ccsdesc[300]{Human-centered computing~HCI theory, concepts and models}

\keywords{Slop, Generative AI, Culture, LLMs, Societal Impact}


\maketitle

\section{Introduction}

The proliferation of AI-generated “slop” is increasingly recognized as an important trend with significant impacts in our creative, information, and cultural economies \cite{heuser_cultural_2025, warzel_tool_2025}. Standing in contrast to the well-marketed promises of AI’s potential for superhuman cognitive processing, a lot of what AI currently offers seems more like the off-cuts of intelligence: a rapidly growing heap of output generated from the same raw material but without the verve, depth, or authenticity of the version made by humans.

AI Slop is often compared to pollution. It is polluting corporations, with previously diligent workers now relying on AI to generate subpar “workslop” which offers the veneer of competence but without the underlying substance \cite{niederhoffer2025ai}. It is polluting culture, with an endless deluge of cheap, pointless content circulating on the internet in place of stuff people might actually care about \cite{madsen2025ai}. But the commonly held view that AI Slop can be understood solely as a scourge crowding out legitimate content leaves an important question unanswered. If AI Slop has so little value, why does it continue to proliferate?

In this paper, we argue that AI Slop should not be dismissed as mere digital detritus, but instead should be taken seriously as an object of study in its own right. This view has two components. The first is that AI Slop has a social function: it offers a supply-side solution to a variety of problems in cultural and economic demand. The second is that AI Slop has aesthetic value: cultural critics have a history of initially dismissing “low” culture only to later admit its significance. Taken together, this perspective presents a case for why slop matters.

\section{AI Slop serves a social function}

People want more content than humans can supply. This does not mean that, taken as a collective whole, there is not enough content in the universe of extant media. Rather, for any given person, there is content they would be willing—even eager—to consume that is not yet being provided. However, making this content comes at a cost. The space of hyper-specialized content people would consume is vast. But for most of this niche content the audience is too small to justify its production. In economics, the balance between the benefit to potential consumers and the cost of production is called optimum product diversity \cite{dixit1977monopolistic}. 


The effect is more general---and potentially more insidious---than simply tailoring content to an individual’s specific interests or the likeness of someone they know. For example, employees can use AI Slop to produce an interminable stream of banal deliverables despite their lack of belief that the work really matters. Dictators can use AI Slop to generate enough propaganda to maintain popular support for their regime. Scientists can use AI Slop to publish a procession of incremental advances. 

Or at least they can try. Humans are exquisitely sensitive to cost-benefit trade-offs in their decision-making \cite{kool_decision_2010, shenhav2017toward, kool2018mental}. This kind of cost-benefit mechanism goes a long way toward explaining the proliferation of AI Slop. On the production side, it offers a kind of near-zero cost means of attempting to achieve one's goals (e.g., generating content that someone, somewhere might want to consume). On the consumption side, it obviates the typical costs associated with finding or generating content that is uniquely suited to one's own specific needs or preferences. The content does not have to be especially rewarding or high quality when the cost of producing or attaining it is so low.

So, what function does AI Slop serve? In our analysis, the answer to this question has to do with how it upends the usual dynamics of optimum product diversity by rewriting cost-benefit tradeoffs for producing content at both an collective economic and individual psychological level. But this is less of an answer than a justification of the question. AI Slop does \textit{something}. What exactly this is depends on the specific use case. We should seek to understand those varied functions, not explain them away.

\section{AI Slop has aesthetic value}

Cultural analysis has often followed the cyclical trend that “low” content is initially dismissed by critics, only to later be appreciated as a form of legitimate---sometimes even important---cultural production \cite{bourdieu_distinction_1984}. For example, in a highly influential essay, Clement Greenberg contrasted the emerging “high” genre of Avant Garde with the “low” genre of cheap, commercial Kitsch \cite{Greenberg1939}. He described Kitsch as the formulaic, mechanical content consumed by the “urban masses” to offer “the diversion that only culture of some sort can provide.” He decried this content as a parasite on “fully matured” culture, drawing its “life blood” from the “reservoir of accumulated experience” and converting it into the “devices, tricks, stratagems, rules of thumb, themes” that form “the debased… simulacra of genuine culture.” It makes for a pretty good description of AI Slop.


However, some of what Greenberg maligned as kitschy is now thought of as having obvious cultural value. Tin Pan Alley songs are referred to as the “Great American Songbook,” while Norman Rockwell has his own museum; both were singled out by Greenberg as quintessentially Kitsch. More generally, Kitsch—along with comparable concepts such as Camp \cite{Sontag1964} and Pastiche \cite{dyer_pastiche_2007}—has been rigorously theorized since Greenberg’s time as its own worthwhile aesthetic genre \cite{calinescu_five_1987, jameson_postmodernism_1993}. Still, some might argue that AI Slop is different: that it is uniquely malignant within humanity’s long history of leveraging scalable, mechanical systems for cheap, popular output. That’s what Greenberg said, too.

While critics have focused on its detrimental effects, we are interested in how this content is used by people to make sense of the world around them and facilitate collective meaning-making \cite{Geertz1973, Williams1977}. This kind of content—whether Kitsch in the 1930s, or AI Slop in the 2020s—is a vehicle for conveying shared sentiment and identity. Contemporary examples might include a dialogues with a simulated pop culture figure, like Lewis Hamilton or a Power Ranger, or a photorealistic image of oneself doing an activity one has never done (e.g., doing an elaborate dance, walking on the moon). For certain people, a Power Ranger might be just the kind of perspective from which they wish they could elicit insight, just as walking on the moon might be an appropriate metaphor for a particular state of mind. 

The term “slop” implies a normative judgment. It is sloppy in the sense that it is aesthetically inferior: a poorly constructed facsimile of the real thing. This assumption writes off possibly that people could use so-called slop as a legitimate site of meaning- and sense-making. And even if AI Slop is somehow inferior, that doesn’t preclude people from finding meaning in it. Humans show an ingenious proclivity for taking whatever is in front of them and finding ways of using it to express what really matters to them \cite{kommers_sense-making_2025, Bruner1990}. If we have any investment in a descriptive understanding of this kind of cultural meaning-making, we should care about AI Slop.

\section{Prototypical AI Slop}

AI Slop has so far resisted formal definition. While recent computational work has sought to automate subjective judgments of AI Slop \cite{shaib_measuring_2025}, it remains in a you-know-it-when-you-see-it stage of conceptual refinement. AI Slop is not a natural kind: its boundaries will not be formally delineated in a way that conclusively separates slop from non-slop. However, it is possible to offer features that most AI Slop has in common, as a kind of family resemblance \cite{rosch1975family}. We propose that AI Slop has the following prototypical properties. It is AI-generated content that is characterized by:

\emph{Superficial Competence.} AI Slop demonstrates a kind of competence that would take real expertise or skill to achieve without AI; this is belied by a lack of underlying substance, craft, or communicative intent. For example, AI-generated work memos have good grammar; even Shrimp Jesus is photorealistic. While this kind of content is based directly on human activity, the wide-scale dissemination of it in such a titrated form creates a new category of AI-generated content. This is characterized by overtly “idealized” output, flatting out individual variation into canonical patterns \cite{heuser_cultural_2025, wright2025epistemic, yang2025alignment}---reflected, for instance, in accusations that liberal use of an em dash is a reliable signature of AI writing. 

\emph{Asymmetric Effort.} AI Slop is generated with a facile prompt and requires little or no effort of the kind that would be necessary to create such an output without AI. However, this does not preclude \textit{any} effortful aspect—for example, trying different prompts to achieve a specific sloppy effect or integration into some larger industrial process. 

\emph{Mass Producibility.} AI Slop is designed to work within a digital ecosystem of production and distribution to reach a mass audience. While many of the outputs may be intended for personalized consumption, they are, crucially, mass produce-\textit{able}.

\section{Dimensions of Variation}

While AI Slop tends to have prototypical family resemblances, the category is also heterogeneous. Workslop is not the same slop that is found on TikTok; it takes different forms in different environments, just as it varies across visual, auditory, and textual media \cite{mcnulty2022content}. With that in mind, we propose three main dimensions of variance:

\emph{Instrumental Utility.} Some AI Slop is generated with a particular purpose in mind. It is sloppy because it tries to accomplish a specific goal while requiring an exertion of significant effort. This is a separate consideration from our argument that much AI Slop can be construed to have some social function. While it is almost always possible to posit a function for AI Slop \textit{post hoc}, that does not mean that it was initially created with such a purpose in mind. 

\emph{Personalization.} Some AI Slop offers a way for content to resemble a specific person’s voice or likeness which previously could not easily be generated. This may result in content that may not especially be notable except to the person who is depicted in it or someone who knows them; for example, while videos of people dancing in their underwear may always have been readily available on the internet, the value of such a video shifts when it is a member of your family who stars in it. 

\emph{Surrealism.} Some AI Slop can have a heightened quality, intensifying otherwise familiar tropes or aesthetics to an unusually large degree \cite{heuser_cultural_2025}. This is related to the “hallucinatory” quality of AI systems. They don't just confabulate plausible material \cite{sui2025critical, sui2024confabulation}; they can continue to double down on the confabulation until it is ludicrously implausible. 

\section{Future Directions}

We argue that academic work should seek to further investigate and refine our understanding of AI Slop, as presented in this paper. This is part of the wider current requiring us to assess the impact of how generative AI systems impact a range of activities, from scientific \cite{evans2025after, holtzman2025prompting} to cultural production \cite{kommers2025computational}. Specifically, we see three general tasks as crucial for a full accounting of AI Slop.

The first is an investigation of its formal properties: What separates AI Slop from non-slop? AI Slop is constituted both by formal features and human judgments; any investigation into defining it will require an understanding of both. For example, scalable empirical experiments and qualitative analysis are needed to delineate what people label as AI Slop and the conditions on which those judgments are based \cite{kommers2025meaning}, while datasets for AI Slop can help to automate the identification of sloppy content or run large scale analysis of its shared properties.

The second is an investigation of its social functions: What do people use AI Slop for? In this paper we offer a claim about the general functions of AI Slop, but there will be much to discover in individual cases. For example, our claim is analogous to an anthropologist saying that religion in general does indeed serve a social function—whereas the bulk of anthropological work looks at specific religions in specific cultural contexts.

The third is an investigation of its cultural and economic impacts: How will the proliferation of AI Slop change the kind of content people want? AI Slop will exert significant influence on how digital platforms are designed in the future, with large AI companies already investing in TikTok-style AI-first platforms such as Sora. The economic effect of this trend is likely to be huge: the slop economy has the potential to be the next attention economy. How the shifting dynamics supply and demand for content—at work, on the internet, in one’s personal life—will be an important subject for further research.

\bibliographystyle{ACM-Reference-Format}
\bibliography{main}

\end{document}